\documentclass[aps,preprint,amsmath,amssymb,nofootinbib]{revtex4}

\usepackage{graphicx}
\usepackage{dcolumn}
\usepackage{bm}
\usepackage[dvipsnames]{xcolor}
\usepackage[utf8]{inputenc}
\usepackage{color}
\usepackage{subfigure}
\usepackage{adjustbox}
\usepackage{float}
\usepackage{amssymb}

\newcommand{\AdS}{\mathrm{AdS}}
\newcommand{\QNM}{\mathrm{QNM}}

\newcommand{\dg}{\mathrm{d}}
\newcommand{\ii}{\mathrm{i}}
\newcommand{\ee}{\mathrm{e}}
\newcommand{\RR}{\mathbb{R}}

\begin{document}

\title{Time-Crystalline Phase in a Single-Band Holographic Superconductor}

\author{Chi-Hsien Tai}
\email{xkp92214@gmail.com}
\affiliation{Department of Physics and Center for Theoretical Physics, Chung Yuan Christian University, Taoyuan, Taiwan}

\author{Wen-Yu Wen}
\email{wenw@cycu.edu.tw}
\affiliation{Department of Physics and Center for Theoretical Physics, Chung Yuan Christian University, Taoyuan, Taiwan}

\begin{abstract}
We investigate the emergence of a time-crystalline phase in a single-band holographic superconductor, extending the the Anti-de Sitter/Conformal Field Theory (AdS/CFT) framework. By incorporating a nonlinear gauge-scalar coupling and external driving, we derive coupled equations of motion for the plasma and Higgs modes, analogous to those in high-$T_c$ superconductors. Multi-scale analysis reveals a sum resonance with subharmonic growth indicating broken time-translation symmetry. We perform numerical computation of quasinormal mode and demonstrate the transition to the time-crystalline phase. The holographic model may serve as a robust tool for studying strongly coupled time crystals.
\end{abstract}                                                                                                         

\maketitle

\section{Introduction}

Time crystals represent a novel phase of matter that spontaneously breaks time-translation symmetry, manifesting as periodic motion in the absence of external periodic driving. This concept, originally proposed as a theoretical curiosity, has evolved into a vibrant area of research bridging condensed matter physics, quantum optics, and non-equilibrium statistical mechanics. The idea was first introduced as an extension of spatial crystals to the time domain, where a system in its ground state exhibits periodic behavior without energy input \cite{Shapere:2012nq, Wilczek:2012jt}. However, early proposals faced challenges related to no-go theorems in equilibrium systems, leading to refinements that emphasized non-equilibrium, driven, or Floquet-engineered setups \cite{Bruno:2013mva, Else:2016yfq}. Over the past decade, time crystals have been realized experimentally in diverse platforms, including trapped ions, ultracold atoms, and spin chains, where discrete time-translation symmetry breaking leads to subharmonic responses stable against perturbations \cite{Else:2019ksr, Giergiel:2020oex}.

In high-Tc superconductors, recent experimental and theoretical advances have uncovered signatures of time-crystalline behavior arising from the nonlinear coupling between collective modes under optical driving \cite{Homann:2020}. Specifically, the interplay between the Higgs mode, corresponding to amplitude fluctuations of the superconducting order parameter, and the Josephson plasma mode, associated with phase oscillations, has been shown to generate subharmonic responses when resonantly driven, a hallmark of discrete time-crystal formation. These phenomena have been observed in materials like cuprates and iron-based superconductors, where ultrafast laser pulses induce coherent oscillations that persist beyond thermal relaxation times \cite{Kleiner:2021, Nevola:2023}. Theoretical models, often based on lattice simulations or effective field theories, highlight how parametric resonances amplify these modes, leading to long-lived periodic states that challenge traditional notions of dissipation in open systems \cite{Ojeda:2021, Collado_2023}. Persistent multi-frequency dynamics of superconducting order parameters are noted in related studies, suggesting potential applications in quantum information processing and non-equilibrium control \cite{Dolgirev:2022}.

A comprehensive dynamical phase diagram of periodically driven BCS superconductors was recently established by Ojeda Collado et al. \cite{Collado_2023}, revealing distinct regimes: Rabi-Higgs, gapless, synchronized-Higgs, and time-crystal, as functions of drive amplitude and frequency. Our holographic construction operates in the strongly coupled regime, where the dual gravitational description captures non-perturbative physics inaccessible to weak-coupling BCS mean-field theory. The emergence of subharmonic responses in our two-mode model through resonant nonlinear coupling between Higgs and plasma modes places it squarely within the time-crystal regime of this broader classification. While the microscopic origin differs (holographic bulk dynamics versus lattice BCS quasiparticles), the effective parametric amplification and synchronization mechanisms share important universal features. This suggests that time-crystalline order under periodic driving may be robust across coupling strengths, with holography providing a controlled theoretical laboratory for the strongly correlated limit.

To model such strongly coupled, driven many-body systems, the framework of holographic duality, as pioneered by Hartnoll et al. \cite{Hartnoll:2008kx}, provides a powerful tool. By leveraging the Anti-de Sitter/Conformal Field Theory (AdS/CFT) correspondence, a gravitational description in the bulk can be mapped onto a strongly interacting boundary theory, here representing a superconducting state. This approach naturally captures emergent phenomena in strongly correlated systems where traditional perturbative methods fail, such as near quantum critical points or in the presence of strong electron-phonon interactions \cite{Huang:2023ihu}. Holography has successfully modeled high-Tc superconductivity features, including pseudogap phases and strange metal behavior, by treating the dual gravity theory as a mean-field approximation to quantum many-body physics.

In this work, we extend the single-band holographic superconductor model to explicitly exhibit time-crystalline dynamics. We introduce a nonlinear interaction term between the scalar (order parameter) and gauge fields, incorporate external optical driving, and derive an effective equation of motion (EOM) for the coupled mode system. Using multi-scale analysis, we estimate sharp resonance conditions where the driven system enters a time-crystalline phase. These analytical predictions are corroborated by extensive numerical simulations of the full nonlinear EOMs, revealing the existence of subharmonic peaks. By bridging lattice-based simulations of high-Tc superconductors with holographic methods, our model offers insights into universal mechanisms of time-crystal formation in strongly coupled systems, potentially guiding future experiments in optically pumped quantum materials \cite{Basov:2017}.

Our results not only confirm the emergence of a driven time crystal in a holographically dual superconductor but also provide a controlled theoretical laboratory for exploring non-equilibrium symmetry breaking in strongly correlated matter. These findings bridge optical control of quantum materials with gravitational descriptions of criticality, offering new insights into the universal features of dynamic phases under periodic driving.

\section{Bulk model and condensed background}

We adopt the probe limit Abelian Higgs action \cite{Hartnoll:2008kx}, which reads
\begin{equation}
S = \int d^4 x \sqrt{-g} \left[ R + \frac{6}{L^2} - \frac{1}{4} F_{\mu\nu} F^{\mu\nu} - |D_\mu \psi|^2 -m^2 |\psi|^2 \right].
\label{eq:action}
\end{equation}

Although the scalar field $\psi$ and $U(1)$ gauge field $A$ behaves like those in the usual Ginzburg-Landau theory, their asymptotic expansion at the boundary encodes physical quantities, such as the condensation, chemical potential and charge density, in the dual field theory.

We work in the planar \(\AdS_4\) black--brane geometry:
\begin{equation}
\dg s^2 = \frac{L^2}{u^2}\Big(-f(u)\,\dg t^2 + \frac{\dg u^2}{f(u)} + \dg x^2 + \dg y^2\Big),\qquad f(u)=1-u^3/u_+^3,
\label{eq:metric}
\end{equation}
with radial coordinate \(u\in(0,u_+)\), where boundary at \(u\to 0\) and future horizon at \(u\to u_+\).  For simplicity, we set radius of curvature $L=1$ and horizon $u_+=1$ throughout the paper.  The Hawking temperature, $T = 3/4\pi$, also characterizes the critical temperature below which condensation occurs in the dual superconductor. 

We take the condensed static background in radial gauge \(A_u{=}0\):
\begin{equation}
A=\phi_0(u)\,\dg t, \qquad \psi=\psi_0(u)\in\RR,
\end{equation}
obeying the coupled ODEs
\begin{align}
\psi_0'' + \Big(\frac{f'}{f} - \frac{2}{u}\Big)\psi_0' + \Big(\frac{q^2\phi_0^2}{f^2} - \frac{m^2}{u^2 f}\Big)\psi_0 &= 0,\label{eq:bg-psi}\\
\phi_0'' - \frac{2}{u}\phi_0' - \frac{2 q^2 \psi_0^2}{u^2 f}\,\phi_0 &= 0.\label{eq:bg-phi}
\end{align}
Regularity at the horizon enforces \(\phi_0(1)=0\), while near the boundary we have the asymptotics
\begin{equation}
\psi_0(u) = \psi^{(1)}u + \psi^{(2)}u^2 + \cdots,\qquad \phi_0(u)=\mu - \rho\,u + \cdots,
\label{eqn_bdc}
\end{equation} 
where we choose $m^2 = -2$ and \(\psi^{(1)}=0\) for a source free condensate in standard quantization.  \(\mu\) and \(\rho\) are chemical potential and charge density.

\begin{figure}[t]
\includegraphics[scale=0.18]{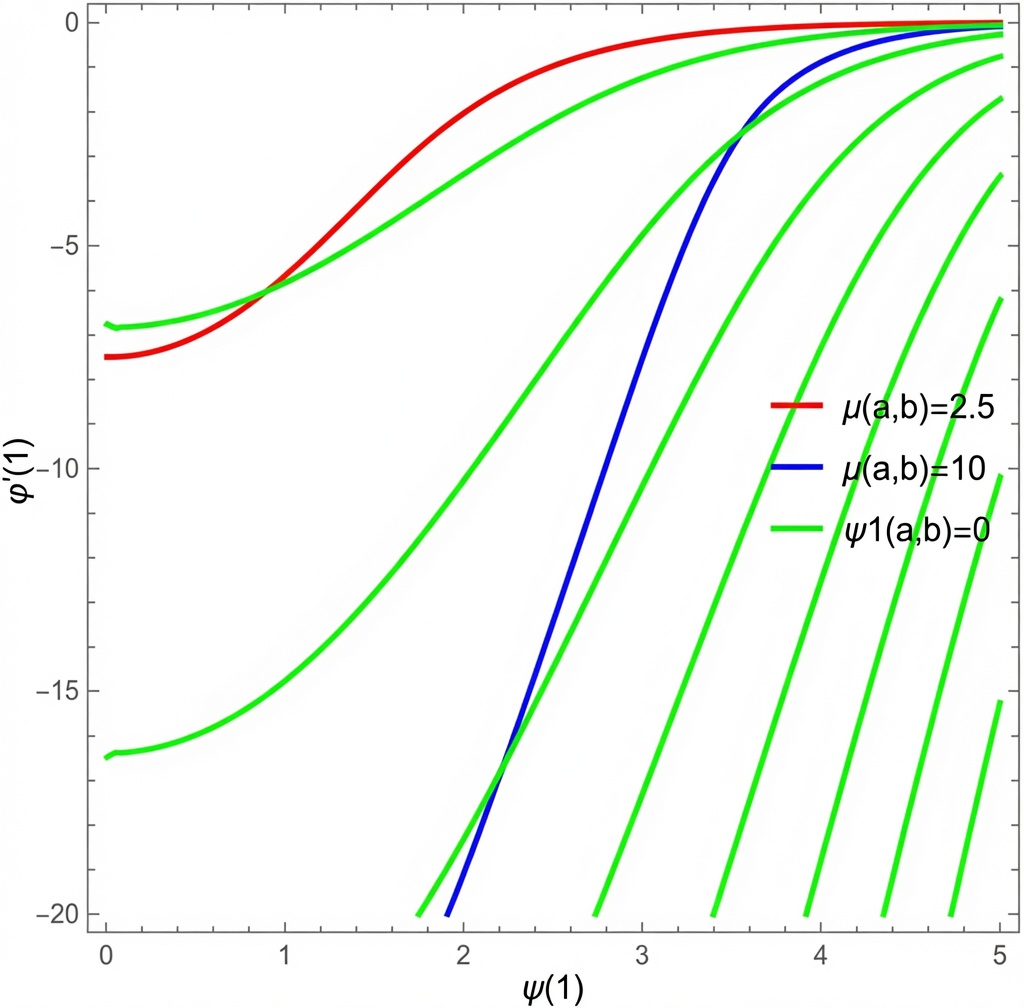}
\includegraphics[scale=0.20]{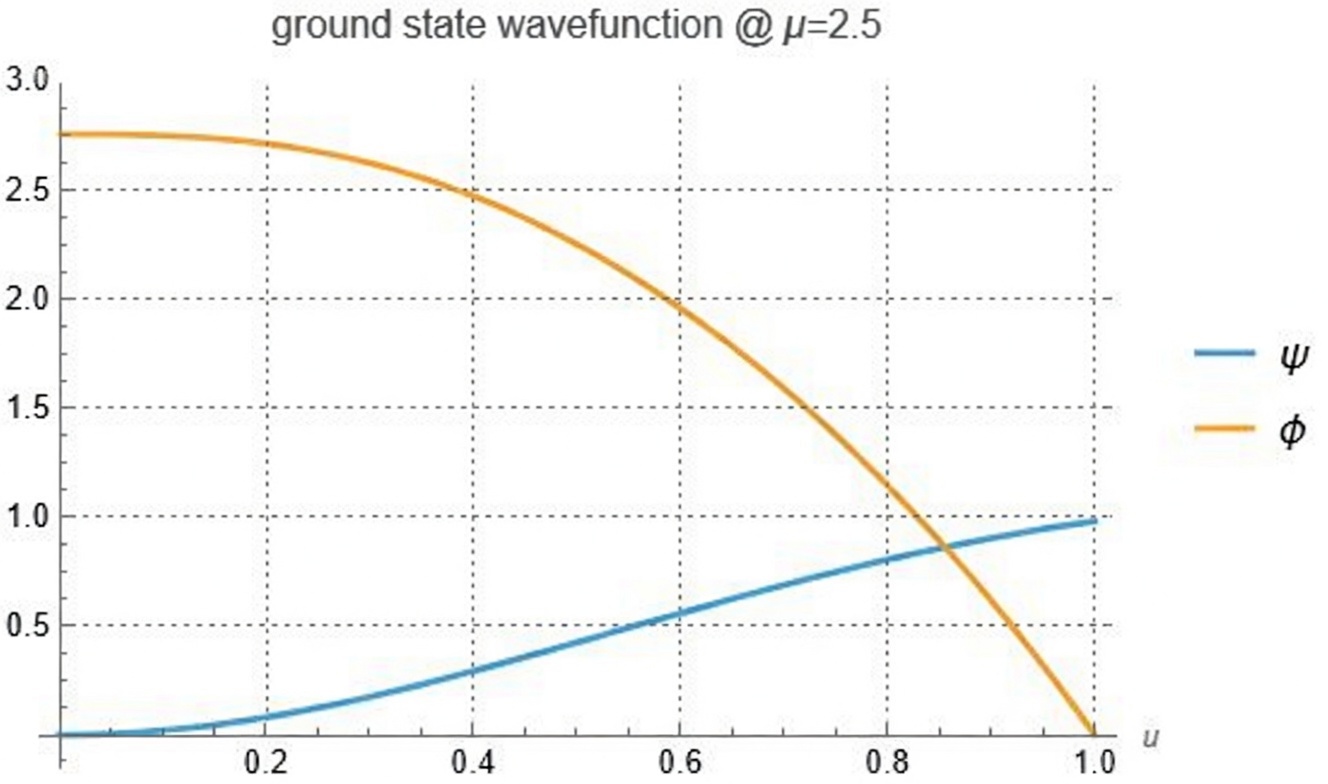}
\caption{\label{fig:gsW}(Left) Contour plot of given boundary conditions for fixed $\mu$ and $\psi^{(1)}=0$ in the shooting parameters space formed by guessing $\psi$ and $\phi^\prime$ at the horizon $u_+=1$.  Solutions exist at intersecting points. For higher chemical potential (for instant, $\mu=10$), there exists multiple solutions for excited states. (Right) In our simulation, we use the ground state wavefunctions $\psi(u)$ and $\phi(u)$ for lower chemical potential, say $\mu=2.5$, to avoid the complexity due to excited modes.}
\end{figure}

\section{Linear excitations and quasinormal modes at \(k=0\)}

G.~Homann et al. \cite{Homann:2020} proposes a light-induced time crystalline state in a high-\(T_c\) superconductor, advancing light control toward non-equilibrium orders and further time crystals in the solid-state domain. It characterizes this state by a time crystal spontaneously breaking time-translation symmetry with subharmonic response.

Termed a Higgs time crystal, it arises from coupled Higgs and Josephson plasma modes in a superconductor with broken \(U(1)\) symmetry and an amplitude oscillation order parameter, as in \(|\psi|^4\) theory. The Higgs mode \(h\) (amplitude fluctuation) and Josephson plasma mode \(a\) (phase oscillation) are orthogonal due to induced electromagnetic interactions. Mapping onto optically driven high-\(T_c\) superconductors as a function of driving frequency \(\omega_d\) and amplitude \(E_0\), non-linear coupling \(\sim a^2 h\) at \(\omega_d = \omega_H + \omega_J\) yields the time crystalline phase, with \(\omega_H\) (Higgs) and \(\omega_J\) (plasma) frequencies. Persistent multi-frequency dynamics of superconducting order parameters are noted in related studies.

In the standard holographic construction, it is sufficient for simple mass term to stabilize the condensate vacuum without introduction of \(|\psi|^4\) term.  This is mainly because scalar fields in the AdS space admits solutions for some \(m^2<0\).  The Higgs mode and plasma mode come naturally from perturbation of gauge field \(\delta A\) and \(\delta \psi\) in the superconducting phase.  The coupling between two modes arises from radial reduction of terms like \(|D\psi|^2\).  The Maxwell on-shell variation provides the boundary term $\int dt\,A_x^{(0)}\,\delta\langle j_x\rangle$, so injects the drive in the reduced description.  At last, our dynamics has dissipation due to ingoing flux at the horizon. In linear response, this corresponds to the imaginary part of quasi-normal mode (QNM).  In the following, we will derive each terms in the desired coupled EOMs from the model of holographic superconductor.

\subsection{Transverse gauge (plasma) sector \(A_x\)}
The quadratic Maxwell Lagrangian density for a homogeneous transverse mode at \(k=0\) is
\begin{equation}
\mathcal L^{(2)}_{J}=\tfrac12\sqrt{-g}\Big[(-g^{tt})g^{xx}(\partial_t A_x)^2-g^{uu}g^{xx}(\partial_u A_x)^2\Big]
-\tfrac12\sqrt{-g}\,\frac{2q^2\psi_0^2}{u^2 f}\,A_x^2.
\end{equation}
Using the metric factors gives
\begin{equation}
\mathcal L^{(2)}_{J}=\tfrac12\frac{(\partial_t A_x)^2}{f}-\tfrac12 f(\partial_u A_x)^2-\tfrac12\frac{2q^2\psi_0^2}{u^2 f}A_x^2.
\end{equation}
Project with \(A_x(u,t)=a_x(t)Y_J(u)\). The time--kinetic term reduces to
\begin{equation}
\int_0^1\!du\,\tfrac12\frac{(\partial_t A_x)^2}{f}=\tfrac12\dot a_x^{\,2}\int_0^1\!du\,\frac{|Y_J|^2}{f}
\equiv \tfrac12\dot a_x^{\,2},
\end{equation}
when we \emph{normalize}
\begin{equation}
\int_0^1\!du\,\mathcal W_J|Y_J|^2=1,\qquad \mathcal W_J(u):=\sqrt{-g}(-g^{tt})g^{xx}=\frac{1}{f(u)}.
\end{equation}
The remaining quadratic potential is
\begin{equation}
-\tfrac12 a_x^2\int_0^1\!du\,\Big[f (Y_J')^2+\frac{2q^2\psi_0^2}{u^2 f}Y_J^2\Big].
\end{equation}
Multiply the QNM equation (\ref{eq:QNM-J}) by \(f\) to obtain the Sturm--Liouville form
\begin{equation}
(f Y_J')'+\Big(\frac{\omega^2}{f}-\frac{2q^2\psi_0^2}{u^2}\Big)Y_J=0.
\end{equation}
Further multiplying by \(Y_J\) and integrating over \(u\), boundary terms vanish by the QNM boundary conditions (ingoing horizon, no source at the boundary), yielding
\begin{equation}
\int_0^1\!du\,\Big[f (Y_J')^2+\frac{2q^2\psi_0^2}{u^2}Y_J^2\Big]=\omega_J^2\int_0^1\!du\,\frac{Y_J^2}{f}.
\end{equation}
With the normalization above, the potential reduces to \(-\tfrac12\omega_J^2 a_x^2\). Hence
\begin{equation}
L^{(2)}_{J}=\tfrac12\dot a_x^{\,2}-\tfrac12\omega_J^2 a_x^2.
\end{equation}

\subsection{Higgs (amplitude) sector \(\eta=\delta\psi\)}
The quadratic scalar--gauge Lagrangian density at \(k=0\) reads
\begin{align}
\mathcal L^{(2)}_{H}&=\tfrac12\sqrt{-g}(-g^{tt})(\partial_t\eta)^2-\tfrac12\sqrt{-g}g^{uu}(\partial_u\eta)^2
-\tfrac12\sqrt{-g}\Big(\frac{m^2}{u^2}+\frac{q^2\phi_0^2}{f}\Big)\eta^2\\
&\quad -\sqrt{-g}\,\frac{2q^2\phi_0\psi_0}{f}\,\eta\,\varphi-\tfrac12\sqrt{-g}\,\frac{2q^2\psi_0^2}{u^2 f}\,\varphi^2,
\end{align}
with \(\varphi=\delta A_t\). Using the metric factors gives the time--kinetic weight
\begin{equation}
\mathcal W_H(u):=\sqrt{-g}(-g^{tt})=\frac{1}{u^2 f(u)}.
\end{equation}
Project with \(\eta(u,t)=h(t)Y_H(u)\). Then
\begin{equation}
\int_0^1\!du\,\tfrac12\,\mathcal W_H\,(\partial_t\eta)^2=\tfrac12\dot h^{\,2}\int_0^1\!du\,\mathcal W_H|Y_H|^2\equiv \tfrac12\dot h^{\,2},
\end{equation}
after normalizing \(\int_0^1 du\,\mathcal W_H|Y_H|^2=1\).

Because \(\eta\) mixes with \(\varphi\), the correct potential follows from the coupled QNM eigen problem. Let \(U=(Y_H,\Phi)^T\) solve the Higgs system with QNM boundary conditions. A similar integration by parts then gives
\begin{equation}
\int_0^1\!du\,\Big\{\frac{f}{u^2}|Y_H'|^2+\frac{1}{u^2}\Big(m^2+\frac{q^2\phi_0^2}{f}\Big)|Y_H|^2+\frac{2q^2\phi_0\psi_0}{u^2 f}\,\Re(Y_H^*\Phi)+\frac{q^2\psi_0^2}{u^2 f}|\Phi|^2\Big\}
=\omega_H^2\int_0^1\!du\,\frac{|Y_H|^2}{u^2 f}.
\end{equation}
With the chosen normalization, the quadratic potential reduces to \(-\tfrac12\omega_H^2 h^2\). Hence
\begin{equation}
L^{(2)}_{H}=\tfrac12\dot h^{\,2}-\tfrac12\omega_H^2 h^2.
\end{equation}

\subsection{Damping from horizon flux}
Linear QNMs decay as $e^{-i\omega t}$ with complex $\omega=\Omega-i\Gamma/2$. The retarded time-domain dynamics is reproduced by including a Rayleigh functional \footnote{We direct readers to appendix A for relation between Rayleigh functional and QNMs.}
\(\mathcal R=\tfrac12\gamma_J\dot a_x^{\,2}+\tfrac12\gamma_H\dot h^{\,2}\) with $\gamma_{J/H}=-2\Im\,\omega_{J/H}^{\QNM}$.  We present QNMs for each mode as follows:

\begin{itemize}
    \item Transverse gauge (plasma) mode\\
With \(\delta A_x(u,t)=\Re\{Y_J(u)\,\ee^{-\ii\omega t}\}\) and other components unperturbed, the linearized Maxwell equation gives
\begin{equation}
Y_J'' + \frac{f'}{f}Y_J' + \Big(\frac{\omega^2}{f^2} - \frac{2 q^2\psi_0^2}{u^2 f}\Big)Y_J = 0.
\label{eq:QNM-J}
\end{equation}
Quasinormal boundary conditions are: ingoing at the horizon, implying \(Y_J\propto (1-u)^{-\ii\omega/(4\pi T)}=(1-u)^{-\ii\omega/3}\), and no source at the boundary, saying \(Y_J\sim u\). The resulting complex eigen frequency is denoted
\begin{equation}
\omega_J^{\QNM}=\Omega_J - \ii\,\Gamma_J/2,
\end{equation}
where \(\omega_J\equiv \Re\,\omega_J^{\QNM}=\Omega_J,\quad \gamma_J\equiv-2\Im\,\omega_J^{\QNM}=\Gamma_J\).
    \item Higgs (amplitude) mode\\
In radial gauge \(\delta A_u=0\) and with real background phase, we consider \(\delta\psi(u,t)=\eta(u)\,\ee^{-\ii\omega t}\) and \(\delta A_t(u,t)=\varphi(u)\,\ee^{-\ii\omega t}\). The coupled linear system is
\begin{align}
\eta''+\Big(\frac{f'}{f}-\frac{2}{u}\Big)\eta' + \Big(\frac{\omega^2}{f^2} + \frac{q^2\phi_0^2}{f^2} - \frac{m^2}{u^2 f}\Big)\eta + \frac{2 q^2\phi_0\psi_0}{f^2}\,\varphi &= 0,\label{eq:QNM-H-eta}\\
\varphi'' - \frac{2}{u}\varphi' - \frac{2q^2\psi_0^2}{u^2 f}\,\varphi - \frac{4 q^2\phi_0\psi_0}{u^2 f}\,\eta &= 0.
\label{eq:QNM-H-phi}
\end{align}
Boundary conditions: ingoing for \(\eta \propto (1-u)^{-\ii\omega/3}\), regular \(\varphi\) at the horizon; at the boundary, no scalar source and \(\varphi(0)=0\) (no chemical potential modulation). The complex eigen frequency
\(\omega_H^{\QNM}=\Omega_H-\ii\,\Gamma_H/2\) defines
\(\omega_H=\Omega_H\) and \(\gamma_H=\Gamma_H\).
\end{itemize}

\subsection{Mode coupling and effective boundary ODEs}
We remark that time-crystalline behavior arises from the nonlinear coupling between collective modes.  In this section, we show that two effective couplings in boundary ODEs can be obtained by dimensional reduction from bulk geometry.
\begin{itemize}
\item{Higgs pumping coefficient $g$ (enters the $h$–equation).}
Recall that we work in radial gauge $A_u=0$ and set
\[
\psi=\psi_0(u)+\eta(u,t),\qquad
A_t=\phi_0(u)+\varphi(u,t),\qquad
A_x(u,t)=a_x(t)\,Y_J(u),\qquad
\eta(u,t)=h(t)\,Y_H(u).
\]
From the bulk scalar kinetic term $-|D\psi|^2$, keeping terms up to \emph{third} order in the small fluctuations $\{\eta,\varphi,A_x\}$ gives the cubic term
\begin{equation}
\mathcal L^{(3)}_{\rm sc}\;=\;
-\,2q^2\sqrt{-g}\,g^{xx}\,\psi_0\,\eta\,A_x^2
\;+\;\cdots .
\label{eq:cubic-from-Dpsi2}
\end{equation}
so projecting with the \emph{Higgs} weight gives
\begin{equation}
g=\frac{2 q^2\displaystyle\int_0^1\!du\,\sqrt{-g}\,g^{xx}\,\psi_0\,Y_H\,|Y_J|^2}
        {\displaystyle\int_0^1\!du\,\sqrt{-g}(-g^{tt})\,|Y_H|^2}.\label{eq:g-overlap}
\end{equation}
This yields the nonlinear drive \(g\,a_x^2\) in
\begin{equation}
\ddot h+\gamma_H\dot h+\omega_H^2 h+\boxed{g\,a_x^2}=0.\label{eq:h_ode}
\end{equation}

\item{Spring–modulation coefficient $\chi$ (enters the $a_x$–equation).}
In the quadratic Maxwell sector, $-|D\psi|^2$ contributes an $A_x$ “mass kernel”
\[
\mathcal M_{A_x}(u)=\frac{2q^2\psi^2}{u^2 f}.
\]
Linearizing the $A_x$ mass kernel with respect to\ the Higgs fluctuation \footnote{We comment that an extra $+\alpha|\psi|^2 A_\mu A^\mu$, if added in the action, can also contribute $+4\alpha\,u^2\psi_0\,\eta\,A_x$.},
\[
\delta (\frac{2q^2\psi^2}{u^2 f})\,A_x^2 =
\Big(\frac{4q^2\psi}{u^2 f}\Big)\delta \psi\, A_x^2=\Big(\frac{4q^2\psi_0}{u^2 f}\Big)\eta\,A_x^2,
\]
and projecting with the \emph{gauge} weight gives
\begin{equation}
\chi=\frac{\displaystyle\int_0^1\!du\,\sqrt{-g}\Big(-\frac{4q^2\psi_0}{u^2 f}\Big)\,Y_H\,|Y_J|^2}
          {\displaystyle\int_0^1\!du\,\sqrt{-g}(-g^{tt})g^{xx}\,|Y_J|^2}.
          \label{eq:chi-overlap}
\end{equation}
It appears as a spring modulation in
\begin{equation}
\ddot a_x+\gamma_J\dot a_x+\big(\omega_J^2+\boxed{\chi\,h}\big)a_x=J\cos(\omega_d t).\label{eq:ax_ode}
\end{equation}
\end{itemize}

We note that $g$ and $\chi$ come from different bulk operators and are projected with different kinetic \emph{weights}. Since those weights are associated with QNMs (dissipation at horizon), the reduction is non-Hermitian; hence they cannot be derived \emph{simultaneously} from single conservative term like $\propto a_x^2 h$ in the effective action. In other words, holographically, they are generically unrelated.  Later, we will show, in principle, they can be individually rescaled.

\section{Multiple-scale analysis and envelopes}

Start with the boundary ODEs collected from previous section:
\begin{align}\label{eqn:BODEs}
\ddot a_x+\gamma_J\dot a_x+\omega_J^2 a_x+\chi\,h\,a_x &= j_0\cos(\omega_d t), \\
\ddot h+\gamma_H\dot h+\omega_H^2 h+g\,a_x^2 &= 0. 
\end{align}
Assume in a weakly driven and coupled regime, we introduce a small bookkeeping parameter
\[
j_0=\varepsilon J,\qquad g=\varepsilon G,\qquad \chi=\varepsilon C,
\]
and define a slow time \(T=\varepsilon t\). We seek following near resonant responses
\[
\omega_d=\omega_J+\varepsilon\sigma_J,\qquad
2\omega_d=\omega_H+\varepsilon\sigma_H^{(2:1)},\qquad
\omega_d=\omega_J+\omega_H+\varepsilon\sigma_H^{(\mathrm{sum})}.
\]
We first assume the fields can be written as fast carrier oscillations with slow envelopes:
\begin{equation}
a_x(t)=A(T)\,e^{i\omega_d t}+\bar A(T)\,e^{-i\omega_d t},\qquad
h(t)=H(T)\,e^{i\Omega t}+\bar H(T)\,e^{-i\Omega t},
\end{equation}
where \(\Omega=\omega_H\) for the $\mathbf{2:1}$ channel and \(\Omega=\omega_d-\omega_J\) for the $\mathbf{sum}$ channel. Let dots denote \(t\)-derivatives and primes denote \(T\)-derivatives.  We remark
\[
\frac{d}{dt}=\partial_t+\varepsilon\partial_T,\qquad
\frac{d^2}{dt^2}=\partial_t^2+2\varepsilon\,\partial_t\partial_T+\mathcal O(\varepsilon^2).
\]

\subsection*{Case I: 2:1 channel, let \(\chi=0\)}
Use \(\Omega=\omega_H\) and keep terms up to \(\mathcal O(\varepsilon)\).  Insert the ansatz in \eqref{eq:ax_ode} with \(j_0=\varepsilon J\). At \(\mathcal O(1)\) we get
\[
(-\omega_d^2+\omega_J^2)\big(Ae^{i\omega_d t}+\bar A e^{-i\omega_d t}\big)=0,
\]
which enforces the near resonance choice \(\omega_d\approx\omega_J\). For \eqref{eq:h_ode}, the \(\mathcal O(1)\) terms vanish identically because \(a_x^2 \sim \mathcal O(1)\) and \(g\sim\mathcal O(\varepsilon)\).

Next collect the coefficients of \(e^{i\omega_d t}\) at \(\mathcal O(\varepsilon)\) from \eqref{eq:ax_ode}:
\[
-2i\omega_d A' - i\gamma_J \omega_d A -2\omega_d\sigma_J A + \frac{J}{2}=0.
\]
Divide by \(2\omega_d\) (replace \(\omega_d\to\omega_J\) at this order) to obtain the \emph{envelope}:
\begin{equation}
A' = -\Big(\tfrac{\gamma_J}{2}+i\sigma_J\Big)A + \frac{J}{4\omega_J}.
\label{eq:Aenv}
\end{equation}

At last, we compute the quadratic drive:
\[
a_x^2 = A^2 e^{2i\omega_d t} + 2|A|^2 + \bar A^{\,2} e^{-2i\omega_d t}.
\]
Insert in \eqref{eq:h_ode}; the term resonant with \(e^{i\Omega t}\) is the one at
\(2\omega_d\approx\omega_H=\Omega\).
Collecting the \(e^{i\Omega t}\) pieces at \(\mathcal O(\varepsilon)\) gives
\[
-2i\Omega H' - i\gamma_H \Omega H -2\Omega\,\sigma_H^{(2:1)} H
 - G\,\frac{A^2}{2}=0.
\]
Divide by \(2\Omega\simeq 2\omega_H\) to obtain
\begin{equation}
H' = -\Big(\tfrac{\gamma_H}{2}+i\sigma_H^{(2:1)}\Big)H
     - i\,\frac{G}{4\omega_H}\,A^2.
\label{eq:HBenv}
\end{equation}
Equations \eqref{eq:Aenv} and \eqref{eq:HBenv} will give the envelope functions.  Other  nonresonant harmonics average out.

\subsection*{Case II: sum channel with back–coupling \(\chi\neq0\)}
Now keep the \(\chi h a_x\) term in \eqref{eq:ax_ode} and choose \(\Omega=\omega_d-\omega_J\). The product \(h a_x\) contains a component
\[
\big(H e^{i\Omega t}\big)\big(\bar A e^{-i\omega_d t}\big)
= (H\bar A)\,e^{-i\omega_J t},
\]
and its complex conjugate. Upon detuning and after removing rapidly oscillating pieces, the term resonant with \(e^{i\omega_d t}\) is \(A\bar H\). Collecting terms similar to those in case I yields
\begin{eqnarray} \label{eq:AenvTriad}
A' &=& -\Big(\tfrac{\gamma_J}{2}+i\sigma_J\Big)A + \frac{J}{4\omega_J}
      - i\,\frac{C}{4\omega_J}\,A\,\bar H, \nonumber\\
H' &=& -\Big(\tfrac{\gamma_H}{2}+i\sigma_H^{(\mathrm{sum})}\Big)H
      - i\,\frac{G}{4\omega_H}\,A^2.
\end{eqnarray}
Again, only the resonant combinations survive; all nonresonant harmonics average out at \(\mathcal O(\varepsilon)\).

We remark that the steady-state linear response of \eqref{eq:Aenv} is
\(
A_\infty=\dfrac{J}{4\omega_J}\dfrac{1}{\gamma_J/2+i\sigma_J},
\)
which, inserted in \eqref{eq:HBenv}, gives the growth rate
\(
\Re\lambda_H=-\gamma_H/2+\dfrac{|G|}{4\omega_H}|A_\infty|^2
\)
and the usual 2:1 onset condition. The growth rate is expected to be negative without driving (decay), but positive with overdriving (blowing up). A similar discussion applies to the sum channel and gives the onset condition:
\begin{equation}\label{eq:sum_onset}
\dfrac{|G||A_\infty|}{2\omega_H}\dfrac{|C||A_\infty|}{4\omega_J} \ge \frac{\gamma_H}{2}\frac{\gamma_J}{2}.
\end{equation}

The back-reaction term proportional to $\chi$ in \eqref{eqn:BODEs} is essential in the sum channel; without it the system exhibits only forced subharmonic response; with it a true instability and a closed Arnold tongue emerge via the feedback loop between the two modes.  We will discuss this in the next section.

\section{Time-crystalline phase from bulk numerics}

In this section we provide a concrete numerical example of time-crystalline phase in the holographic superconductor.  We outline the numerical procedure as follows: 

1. Solve coupled ODEs \eqref{eq:bg-psi} and \eqref{eq:bg-phi} for \(\{\psi_0,\phi_0\}\) with boundary conditions \eqref{eqn_bdc}, given a constant chemical potential. To achieve this, we employ a shooting method from the horizon to the boundary. For stability, we tune the initial condensate value to match the source-free boundary condition \(\psi^{(1)}=0\), achieving convergence to within $10^{-4}$ relative error, see Fig. \ref{fig:gsW}.

2. Compute the lowest QNMs: \(\omega_J^{\QNM},\omega_H^{\QNM}\) via horizon-to-boundary shooting. Define \(\omega_{J/H}=\Re\omega^{\QNM} \), \(\gamma_{J/H}=-2\Im\omega^{\QNM}\). For our pick \(\mu=2.5\), one obtains \(\omega_H^{\QNM} \approx0.951 - i1.676\), \(\omega_J^{\QNM} \approx1.633 - i1.304\).

3. Evaluate \(g\) and \(\chi\) using overlaps by normalizing with the weights \(Y_J\) and \(Y_H\), yielding g\(\approx\) 0.4876 and \(\chi\approx\) 557.7 for the chosen parameters.

4. Insert parameters into the ODEs and solve time-domain evolution by scanning $j_0$ and $\omega_d$; verify the subharmonic peak in the spectrum of h.

We remark that the couplings $g$ and $\chi$ obtained from step $3$ may not be unique but enjoy a scaling symmetry:
\begin{equation}
g \to \lambda_1 g, \quad \chi \to \lambda_2 \chi.
\end{equation}
Then the boundary ODEs \eqref{eq:h_ode} and \eqref{eq:ax_ode} remain the same under the field redefinition:
\begin{equation}
a \to a/\sqrt{\lambda_1\lambda_2}, \quad J \to J/\sqrt{\lambda_1\lambda_2}, \quad h \to h/ \lambda_2.
\end{equation}

\begin{figure}[t]
\includegraphics[scale=0.55]{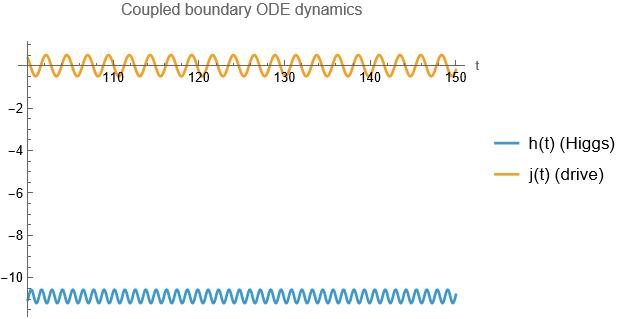}
\includegraphics[scale=0.55]{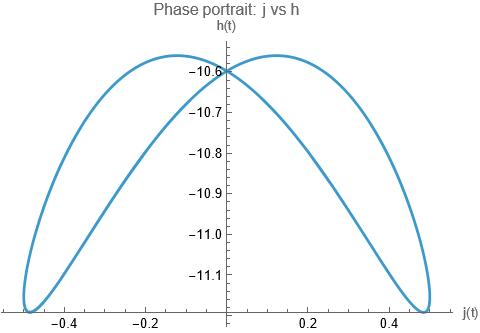}
\includegraphics[scale=0.55]{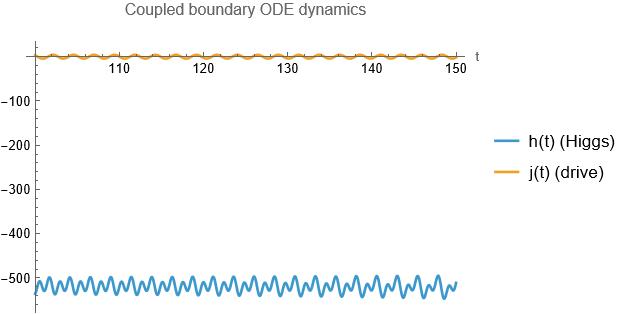}
\includegraphics[scale=0.55]{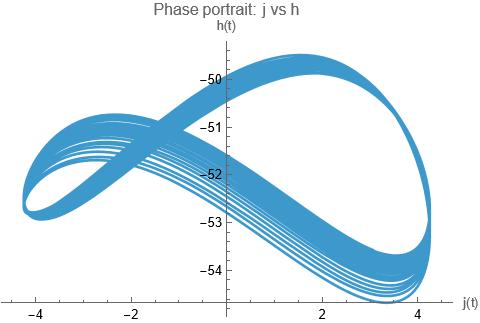}
\caption{\label{fig:wJ} (Top Left) A typical Higgs mode $h$ responds to small driving current $j$ in sum channel.  (Top Right) Phase diagram of $j$ v.s. $h$  shows a closed and stable trajectory. (Bottom Left) Distorted Higgs mode $h$ responds to large driving current $j$ in sum channel, where envelope amplitude is growing and will blow up at later time (not shown), indicating failure of perturbation approach. (Bottom Right) Phase diagram shows uneven and non repeating trajectory.  We remark that the amplitudes of $j(t)$ and $h(t)$ have been rescaled for presentation purpose.  We only show first $150$ sampling points in time domain.}
\end{figure}

\begin{figure}[t]
\includegraphics[scale=0.55]{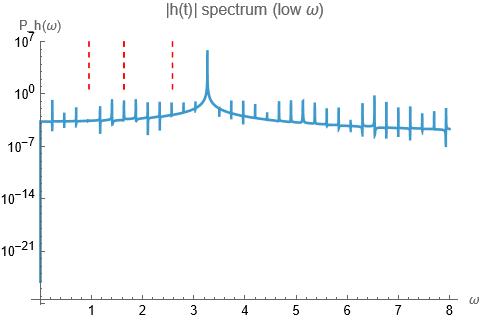}
\includegraphics[scale=0.55]{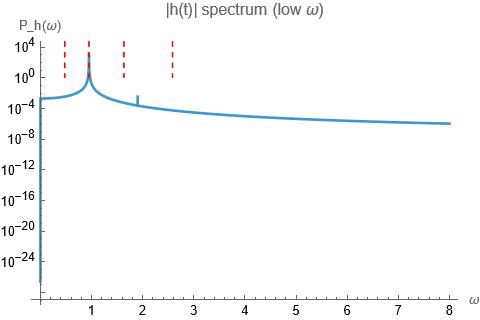}
\includegraphics[scale=0.55]{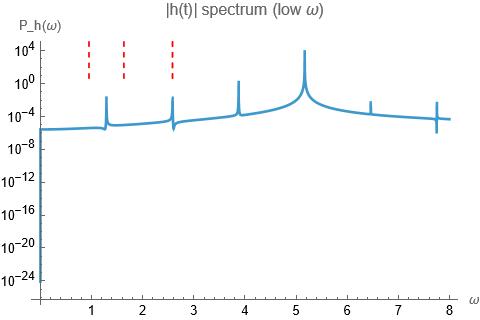}
\includegraphics[scale=0.55]{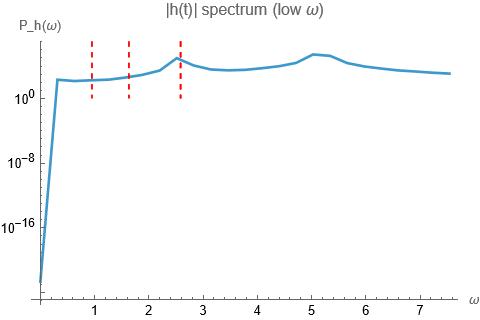}
\caption{\label{fig:wJ2} (Top Left) Power spectrum of Higgs mode in 2:1 channel for $\omega_d = \omega_J$. The vertical red dashed lines indicate $\omega_H$, $\omega_J$, $\omega_H+\omega_J$ from left to right.  Among many resonance peaks, the strongest resonance happens at $2\omega_J$. (Top Right) Power spectrum of Higgs mode in 2:1 channel for $2\omega_d = \omega_H$. The extra vertical dashed line (most left) indicates $\omega_d$. The strongest resonance happens at $\omega_H$. (Bottom Left) Power spectrum of Higgs mode in sum channel for $\omega_d = \omega_J+\omega_H$. Resonance happens at integer multiples of $(\omega_J+\omega_H)/2$. (Bottom Right) Power spectrum of Higgs mode in sum channel, but for runaway Higgs amplitude (large driving current). Vanishing subharmonic peaks signify the system has left time-crystalline phase. We note that parameters for this simulation are $\mu=2.5$, $\omega_H=0.95$, $\omega_J=1.63$, $\gamma_H=3.35$, $\gamma_J=2.61$, $g=4.88$, $\chi=5.58$ (rescaled).}
\end{figure}

Although the nonlinear couplings obtained from the bulk overlap integrals are not unique and enjoy the scaling symmetry. The physically relevant information resides in the dimensionless invariant
\begin{equation}
I = \frac{|GC||A_{\infty}|^{2}}{2\gamma_{J}\gamma_{H}\omega_{J}\omega_{H}}
= \frac{|g\chi|J^{2}}{8\gamma_{J}^{3}\omega_{J}^{3}\gamma_{H}\omega_{H}},
\label{eq:invariant}
\end{equation}
which can be inferred from \eqref{eq:sum_onset}. Linear stability analysis of the slow-envelope equations shows that the $H=0$ state loses stability (onset of a stable subharmonic response) precisely when $I\ge 1$. Because $I$ is invariant under the rescaling, its value computed with the microscopically determined holographic couplings directly measures the distance to the time-crystalline threshold, independent of the arbitrary choice of rescaling used for numerical presentation. For the drive strengths employed in Figs.~2 and 3 we find $I\gtrsim 1$ precisely in the window where clean subharmonic peaks are observed.  As shown in Fig. \ref{fig:wJ}, we found that for small driving current, the Higgs mode oscillates at twice the driving frequency (harmonic response). For large driving, nonlinearity leads to envelope growth and distortion. In Fig. \ref{fig:wJ2}, a finite number of resonant peaks are observed in the spectra of both the 2:1 and sum channels. These subharmonic peaks, in particular, signify the presence of a time-crystalline phase \cite{Greilich:2024}.  At very large driving amplitudes, the Higgs envelope exhibits runaway growth. This behavior occurs well beyond the upper boundary of the Arnold tongue obtained from the two-mode truncation. We interpret it as a signature that the system has left the regime of validity of the reduced model, likely due to the excitation of higher quasinormal modes or strong nonlinear back-reaction not captured by the lowest-mode Galerkin projection.

To establish that the subharmonic response occupies a finite-measure region in parameter space and verify the robustness of the time-crystalline phase with respect to the background, we compare the Arnold tongues for three different chemical potentials, $\mu = 2.3$, $2.5$, and $4.0$, expressed in terms of the dimensionless invariant
\begin{equation}\label{eq:j_crit}
I_{\rm crit}(\sigma_J) = \frac{|g\chi|J_{\rm crit}^2}{8\gamma_J^3\omega_J^3\gamma_H\omega_H},
\end{equation}
where 
\begin{equation}
J_{\rm crit}(\sigma_J) = 2\omega_J\gamma_J\sqrt{\frac{2\gamma_J\gamma_H\omega_J\omega_H}{|g\chi|}\left[\left(\frac{\gamma_J}{2}\right)^2 + \sigma_J^2\right]}.\nonumber
\end{equation}
This quantity is invariant under the rescaling symmetry of the boundary equations and directly measures the distance to the instability threshold. As shown in Fig.~4, the lower boundary of the tongue is universally given by $I=1$ at zero detuning for all three values of $\mu$, while the upper boundary is determined by numerical continuation and corresponds to a saddle-node bifurcation. In all cases a finite, bounded region with $I>1$ exists in which a stable, phase-locked subharmonic response with constant envelope is supported. The persistence of this structure across different background chemical potentials demonstrates that the emergence of the time-crystalline phase is a robust feature of the driven holographic superconductor within the probe limit, rather than an artifact associated with a particular value of $\mu$. We explain the  computation steps in Appendix B.

\begin{figure}[htbp]
\centering
\includegraphics[width=0.92\textwidth]{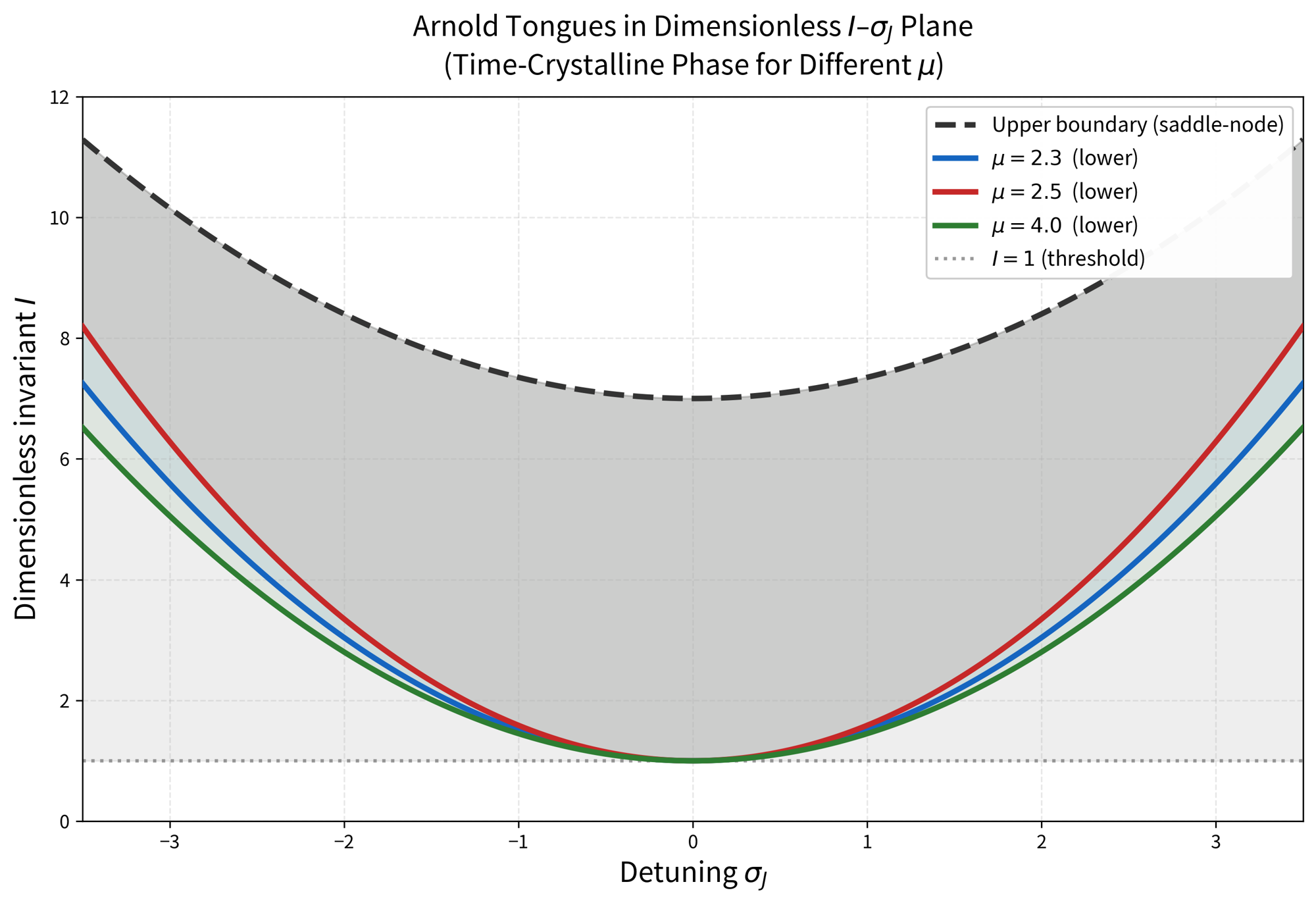}
\caption{Arnold tongues in the dimensionless $I$--$\sigma_J$ plane for three different chemical potentials. The lower boundaries (solid colored curves) are obtained analytically from the linear stability condition of the $H=0$ state and depend on $\mu$ through the damping rate $\gamma_J$ ~\eqref{eq:j_crit}. The upper boundary (thick dashed curve) is determined by numerical continuation of the nonlinear fixed-point equations and corresponds to a saddle-node bifurcation; it is sketched as a common curve for simplicity, but its precise location could vary by different $\mu$ and numerical methods.  We refer the readers to Appendix B for detail. The gray dotted line marks the universal threshold $I=1$. The existence of a finite region of stable time-crystalline behavior between the lower and upper boundaries for all three values of $\mu$ demonstrates the robustness of the phenomenon.}
\label{fig:arnold_tongue_I}
\end{figure}

\section{Discussion and Conclusion}

The projection onto the lowest quasinormal modes (QNMs) provides a controlled one-pole truncation of the bulk dynamics. This approximation is quantitatively reliable in the vicinity of the first plasma resonance and for moderate driving strengths. For stronger driving or when operating away from primary resonances, higher QNMs and couplings among different modes become increasingly relevant. In such regimes, the present framework can be systematically extended by enlarging the QNM basis and repeating the Galerkin reduction, or by reverting to fully backreacted numerical solutions \cite{Hartnoll:2008vx}. We remark on a recent work by \cite{Yang:2023dvk} in which a driven-dissipative
superfluid model using fully backreacted holographic action exhibits a space-time supersolid (STS) phase. Our work is different in the following aspects: (a) The driving in \cite{Yang:2023dvk} is enforced by a time-modulated and spatially inhomogeneous chemical potential, but our driving is by an explicit boundary source; (b) The STS phase in \cite{Yang:2023dvk} appears in strong driving and simultaneously breaks continuous spatial translation, discrete time translation, and U(1); our work, on the other hand, offers a cleaner benchmark for Higgs time crystal physics where the reduced $0+1$ system demonstrates that spatial inhomogeneity is not required for time-crystalline order in holographic superconductors. In our simulation we have ignored the phases occured in \eqref{eq:g-overlap} and \eqref{eq:chi-overlap} for simplicity.  The imaginary parts of $g$ and $\chi$ effectively renormalize the linear damping coefficients and can be absorbed into $\gamma_{J}$ and $\gamma_{H}$ at leading order. The derivation also straightforwardly adapts to alternative scalar quantizations by interchanging source and vacuum expectation value near the AdS boundary.

Building on this controlled reduction, we have presented a self-contained derivation of boundary effective ordinary differential equations governing driven, homogeneous dynamics of a single-band holographic superconductor, with all coefficients fixed from first principles and appropriate rescalings. A multiple-scale analysis of these equations predicts robust subharmonic (time-crystalline) responses, in quantitative agreement with numerical simulations. The formalism is readily generalizable to multi-band holographic superconductors, as well as to inhomogeneous or finite-size drive geometries. Extensions incorporating momentum-dependent couplings, finite-size effects, or more realistic lattice structures can be implemented within the same framework \cite{Wen:2013ufa,Zhang:2025hkb}.

Beyond the minimal setup considered here, several physically motivated directions merit further investigation. Finite-temperature extensions could clarify crossover behavior and the fate of time-crystalline responses near criticality\footnote{We thank \cite{Jeong:2023las} for pointing out their holographic approach on Anderson-Higgs mechanism, where a crossover of Higgs mode was observed at about half $T_c$.  We do not expect to observe this crossover since we have chosen the fixing chemical potential such that boundary temperature is not far away from $T_c$.}, while coupling to competing orders---such as charge density waves or nematic phases---may generate richer dynamical phase diagrams. Such studies could establish connections to experimentally observed nonequilibrium phenomena in bilayer graphene, moiré materials, and related strongly correlated systems \cite{Zhao:2023}. These extensions may also shed light on the interplay between driven superconductivity and other collective modes in complex quantum materials.

More broadly, by applying holographic methods to time-crystalline behavior previously observed in simulated lattice high-$T_c$ superconductors, this work highlights the utility of gauge/gravity duality in capturing intrinsically nonperturbative dynamics inaccessible to weak-coupling approaches. The predicted subharmonic responses closely resemble the experimental signatures observed in optically pumped cuprates, suggesting that holographic models can serve as effective simulators for ultrafast spectroscopy. Future directions include incorporating quantum noise through stochastic holography to assess stability against decoherence, as well as explicitly modeling coupling to external baths to realize open-system time crystals. Ultimately, these insights may inform the design of quantum devices that exploit broken time-translation symmetry to enhance coherence times or enable novel Floquet engineering protocols.

At last, we remark that in the holographic framework the external optical drive enters through the boundary value of the gauge field, which sources a $U(1)$ current in the dual boundary theory. The conjugate response is the expectation value of this current, directly related to the electromagnetic vector potential and electric field at the boundary. The Higgs mode $h(t)$ modulates the condensate amplitude and thereby the charge density; this in turn affects the plasma frequency and the frequency-dependent conductivity. Consequently, persistent subharmonic oscillations of $h(t)$ are expected to imprint observable subharmonic sidebands or higher-harmonic generation features onto the nonlinear THz response, e.g., in differential transmission $\Delta E/E$ or pump-probe reflectivity spectra. Although a full computation of the driven conductivity tensor lies beyond the present two-mode truncation, the reduced model already captures the essential frequency content that would appear in such measurements.

\begin{acknowledgments}
Early work of this project has been presented in the NSYSU (Taiwan) and we thank Prof. Dimitrios Giataganas and Prof. Chia-Yi Ju for inspiring discussions. This work was supported in part by Taiwan’s Ministry of Science and Technology under grant numbers NSTC 112-2112-M-033-003-MY3, NSTC 114-2811-M-033-004 and NSTC 114-2112-M-033-011, and by the National Center for Theoretical Sciences (NCTS). The authors acknowledge the use of the ChatGPT and Grok for assistance with language editing, code formatting, and manuscript organization.
All scientific content, derivations, and conclusions were developed and
verified by the authors. Numerical calculations were performed using \textsc{Mathematica}, with additional analysis carried out using custom Python scripts.
\end{acknowledgments}

\appendix

\section{Damping from horizon flux and Rayleigh functional}
\label{app:horizon-flux-derivation}
This appendix shows explicitly why the dissipative terms in the reduced ODEs can be encoded by a Rayleigh functional
$\mathcal R=\tfrac12\gamma_J\dot a_x^{\,2}+\tfrac12\gamma_H\dot h^{\,2}$ with
$\gamma=-2\,\Im\omega^{\QNM}$ for each channel.

\subsection*{a. Energy balance and radial flux}
For a quadratic field theory on a fixed black–brane background, the canonical energy of a homogeneous perturbation $\Phi$ is
\begin{equation}
 E[\Phi]=\frac12\int_0^1\!du\,\Big(W\,|\partial_t\Phi|^2+P\,|\partial_u\Phi|^2+V\,|\Phi|^2\Big),
\end{equation}
with time–kinetic weight $W(u)=\sqrt{-g}(-g^{tt})\times(\text{index factor})$.
Varying the action and using the equations of motion yields the balance law
\begin{equation}
 \frac{dE}{dt}= -\Big[\mathcal F_{\rm rad}\Big]_{u=0}^{u=1},
\end{equation}
where $\mathcal F_{\rm rad}$ is the radial energy flux density. For QNM boundary conditions, the boundary source vanishes at $u\to0$, and hence at $dE/dt=-\mathcal F_{\rm hor}\le0$.
Near the horizon, use the Eddington–Finkelstein (EF) time $v=t+\!\int du/f$. A regular mode behaves as $\Phi\sim e^{-i\omega v}$, which implies a positive ingoing flux $\mathcal F_{\rm hor}>0$ whenever $\Re\omega>0$.

\subsection*{b. QNM Wronskian identity and $\gamma=-2\,\Im\omega$}
Write the radial equation in Sturm–Liouville form,
\begin{equation}
 (pY')'+(\lambda w-q)Y=0,\qquad \lambda=\omega^2,
\end{equation}
with $(p,w)=(f,1/f)$ for $A_x$ and $(p,w)=(f/u^2,1/(u^2f))$ for the Higgs scalar.
Multiply by $Y^*$, subtract the conjugate and integrate:
\begin{equation}
 \Big[p\,(Y^*Y'-Y Y'^*)\Big]_{0}^{1}+2i\,\Im(\lambda)\int_0^1\!du\,w\,|Y|^2=0.
\end{equation}
At $u=0$ the surface term vanishes due to the boundary condition without source. Using the EF behavior of the initiating $Y\sim(1-u)^{-i\omega/3}$ and $f\simeq3(1-u)$, the horizon term is evaluated to $-2i\,\Re\omega\,|Y(1)|^2$. For small damping, $\Im(\lambda)=2\Re\omega\,\Im\omega$, hence
\begin{equation}
 -\Im\omega\int_0^1\!du\,w\,|Y|^2=\tfrac12|Y(1)|^2.
\end{equation}
With the kinetic normalization $\int w|Y|^2=1$, this gives $-\Im\omega=|Y(1)|^2/2$ and therefore
\begin{equation}
\gamma\;:=\;-2\,\Im\omega^{\QNM}=|Y(1)|^2.
\end{equation}

\subsection*{c. Matching to Rayleigh dissipation}
For a reduced coordinate $q(t)$ with $L=\tfrac12\dot q^2-\tfrac12\omega^2 q^2$, Rayleigh’s functional
$\mathcal R=\tfrac12\gamma\dot q^2$ produces $\ddot q+\gamma\dot q+\omega^2 q=0$ and power loss
$P_{\rm diss}=-\gamma\dot q^2=-2\mathcal R$. Using the mode truncation $\Phi(u,t)=q(t)Y(u)$ and normalization 
$\int w|Y|^2=1$, the bulk flux identity implies $\mathcal F_{\rm hor}=\gamma\dot q^2$ with $\gamma=-2\Im\omega$.
Hence, the Rayleigh functional in the reduced ODEs exactly reproduces the horizon energy sink implied by the QNM imaginary parts.

\section{Numerical Continuation and Detection of the Saddle-Node Boundary}

This appendix provides the numerical procedure over a range of detunings $\sigma_J$ to yield the
upper boundary of the Arnold tongue.

\subsection*{a. Fixed-point equations of the slow-envelope system}

In the sum channel, the time evolution of the slow envelopes \(A(T)\) and \(H(T)\) is governed by \eqref{eq:AenvTriad}. Stationary states satisfy the nonlinear algebraic system:
\begin{align}
0 &= -\Bigl(\frac{\gamma_J}{2} + i\sigma_J\Bigr)A + \frac{J}{4\omega_J} - i\frac{C}{4\omega_J} A \bar{H}, \label{eq:FP_A} \\
0 &= -\Bigl(\frac{\gamma_H}{2} + i\sigma_H\Bigr)H - i\frac{C}{4\omega_H} A^2. \label{eq:FP_H}
\end{align}
These equations determine both the trivial branch \(H=0\) and the nontrivial time-crystalline branch with finite \(|H|\).

To remove the rotational symmetry, we fix the global phase so that \(A\) is real (\(A = x \in \mathbb{R}\)). Writing \(H = u_r + i u_i\), Eqs.~(\ref{eq:FP_A}) and (\ref{eq:FP_H}) reduce to a system of four real equations,
\begin{equation}
\mathbf{F}(\mathbf{x}; J, \sigma_J) = 0,
\end{equation}
where \(\mathbf{x} = (x, u_r, u_i)\) and the explicit components of \(\mathbf{F}\) are obtained by separating real and imaginary parts.

\subsection*{b. Linear stability and the lower boundary}

The trivial solution \(\mathbf{x}_0 = (A_\infty, 0, 0)\) with
\begin{equation}
A_\infty = \frac{J/4\omega_J}{\gamma_J/2 + i\sigma_J}
\end{equation}
is linearly stable provided all eigenvalues of the Jacobian matrix
\begin{equation}
J(\mathbf{x}_0; J, \sigma_J) = \frac{\partial\mathbf{F}}{\partial\mathbf{x}}\Big|_{\mathbf{x}_0}
\end{equation}
have negative real parts. Setting the real part of the critical eigenvalue to zero yields the analytic expression for the lower boundary of the Arnold tongue given in Fig. \ref{fig:arnold_tongue_I}.

\subsection*{c. Numerical continuation to the upper boundary}

For drive strengths above the lower boundary, a branch of nontrivial solutions with finite \(|H|\) exists. Starting from a known solution \(\mathbf{x}^{(n)}\) at drive strength \(J_n\), we advance to \(J_{n+1} = J_n + \Delta J\) by using \(\mathbf{x}^{(n)}\) as the initial guess for a Newton solver applied to the nonlinear system \(\mathbf{F}(\mathbf{x}; J_{n+1}, \sigma_J) = 0\). The step size \(\Delta J\) is chosen adaptively to maintain convergence. This predictor-corrector procedure traces the entire stable finite-\(|H|\) branch continuously as \(J\) is increased.

The upper boundary of the time-crystalline region occurs at a saddle-node bifurcation, where a stable and an unstable fixed point collide and annihilate. At such a point \((\mathbf{x}_*, J_*)\) the following two conditions are satisfied simultaneously:
\begin{align}
\mathbf{F}(\mathbf{x}_*; J_*, \sigma_J) &= 0, \label{eq:SN1} \\
\det\Bigl( \frac{\partial\mathbf{F}}{\partial\mathbf{x}} \Big|_{\mathbf{x}_*, J_*} \Bigr) &= 0. \label{eq:SN2}
\end{align}
Equation~(\ref{eq:SN2}) states that the Jacobian matrix possesses a zero eigenvalue. In practice we monitor the smallest real eigenvalue of the Jacobian while continuing in \(J\). The value of \(J\) at which this eigenvalue crosses zero (while a solution to Eq.~(\ref{eq:SN1}) still exists) is recorded as the saddle-node point. Repeating this procedure over a range of detunings \(\sigma_J\) yields the upper boundary of the Arnold tongue.

This numerical procedure closes the Arnold tongue and demonstrates that the region of stable, phase-locked subharmonic oscillations with constant envelope is bounded in both detuning and drive strength.

\end{document}